\documentclass[12pt]{iopart}

\begin{document}
\title{Streamers, sprites, leaders, lightning: from micro- to macroscales}
\author{Ute Ebert$^{1,2}$ and Davis D. Sentman$^3$}
\address{$^1$~CWI, P.O. Box 94079, 1090 GB Amsterdam, The Netherlands,
\\$^2$~Department of Physics, Eindhoven University of Technology, The Netherlands,
\\$^3$~Geophysical Institute, University of Alaska, Fairbanks, Alaska}

\begin{abstract}
"Streamers, sprites, leaders, lightning: from micro- to macroscales" was the theme of a workshop in October
2007 in Leiden, The Netherlands; it brought researchers from plasma physics, electrical engineering and
industry, geophysics and space physics, computational science and nonlinear dynamics together around the
common topic of generation, structure and products of streamer-like electric breakdown. The present cluster
issue collects relevant articles within this area; most of them were presented during the workshop. We here
briefly discuss the research questions and very shortly review the papers in the cluster issue, and we also
refer to a few recent papers in other journals.
\end{abstract}

\submitto{Editorial introduction for cluster issue on ``Streamers, Sprites and Lightning'' in \JPD}

\maketitle

\section{Streamers: questions and methods}

\subsection{The pivotal role of streamers in discharges in matter}

Rapid electrical breakdown is a generic and ubiquitous process in nature and technology; its earliest stage
is governed by the formation of filamentary streamers. They occur in lightning in the earth's atmosphere and
on other planets, they are developed for and used in various technical and industrial applications, and they
are studied as a fundamental phenomenon in air, pure nitrogen, argon, and other gasses.  They occur equally
in water, oil, semiconductors and other materials. Streamers play a pivotal role in the complex sequence of
events that start with electron avalanche and culminate in leader or arc discharges, and their internal
processes are the most nonthermal within this sequence.  While the various stages of electrical breakdown
are well known, streamers are intrinsically the most difficult to investigate both experimentally and
computationally, as they are very fast, very far removed from thermal equilibrium and sometimes even from
equilibrium with the local electric field, and as they evolve on multiple scales in both space and time. The
spatial scales range from the collisional mean free path of electrons, through the generic inner structure
of a single streamer with its characteristic field enhancement, and on to the growth of interacting streamer
trees consisting of tens-of-thousands of channels.  The temporal scales range from electron impact collision
times to the characteristic times for the formation of arcs, lightning and afterglow. An understanding of
the full sequence of events in sparks or lightning must therefore include a detailed description of the
streamer. Furthermore, the streamer by itself is active in corona reactors as well as in sprite discharges.
Although various approximate schemes have been proposed for describing streamers, a detailed {\it ab initio}
kinetic description at all spatial and temporal scales does not presently exist.

\subsection{Pertinent questions}

There are a number of persistent theoretical questions concerning the nonlinear dynamical properties of
streamers that have continued to elude fundamental description at the kinetic level, despite many decades of
research. Among these are the necessary and sufficient conditions for their initiation, what determines
their diameters and propagation speeds, when do streamers branch, how do they interact with one another or
with dielectric walls, or with dust or water droplets or ice particles, when do they extinguish,  how do
streamers evolve across strong material or electrical gradients in the underlying medium, how do boundary
conditions related to various kinds of electrodes affect the properties of streamers, whether these be solid
laboratory electrodes or the ionosphere,  how does the gas composition, its electronegativity and
photoionizability and the level of background ionization, determine streamer structure and dynamics, and how
are streamer properties affected by an anisotropic conductivity such as induced by a strong external
magnetic field?

All these questions have been grappled with in the past in one form or another by the various research
groups in their respective laboratory and geophysical investigations, but ad hoc approaches have often had
to be used in the absence of a fundamental theory to provide key parameters.  Examples of this approach are
lumped circuit analogies to determine the current flow in the streamer body behind the head, or 1.5D
computational models for streamers with pre-determined radius, or the dielectric breakdown model to describe
the overall branched structure of many streamers. These have proven to be very useful concepts that embody
considerable physical insight, but a complete streamer theory would be able to provide detailed descriptions
of all these quantities, as well as all other quantities associated with a streamer, from first principles.
It is the development of such a comprehensive theory that stands as one of the goals of streamer research.

\subsection{Methodological progress and the topics of the workshop}

In recent years, research methodology has progressed significantly as computational power and sophistication
of algorithms has advanced, and as faster and more sensitive detection technology has become available: in 
plasma diagnostics and plasma modeling, in observations and modeling of atmospheric electricity, in electric 
pulsed power technology, and in multi-scale computing and nonlinear dynamics. The commonality and complexity 
of the breakdown problem is an incentive to cross disciplinary borders between plasma physics and chemistry,
electrical engineering, industrial applications and geophysics, nonlinear dynamics and computational
science.  In fact, the frequency of cross-referencing among these disciplines in published research papers
has increased considerably in recent years.

The workshop at the Lorentz Center in Leiden, The Netherlands, in October 2007 created a forum for direct
knowledge transfer among the various disciplines, and also for further development of methods and concepts
within each discipline. It focused on the following topics relevant for laboratory, industrial and
atmospheric discharges:

(a)  Microscopic mechanisms in different gases, their experimental identification and model implementation.
This included photoionization, background ionization, electron impact excitation and ionization, thermal and
nonthermal attachment and detachment processes, inelastic collisions, etc. That the rates of the fast 
processes in the streamer head scale to good approximation with gas density, is a key to understanding
why streamers at standard temperature and pressure and sprite discharges at high altitudes of the atmosphere
are physically similar.

(b)  Macroscopic spatial structures.  This includes observation and modeling of single streamers, their
branching or extinction, and interactions among streamers, experimental methods for discharge observations
in the laboratory and the atmosphere, and computational methods for studying their multi-scale nature.

(c)  The electron energy distribution at the streamer tip and inside the discharge and the subsequent plasma
chemistry.  Here, we include plasma chemistry in corona reactors and other industrial applications, the
impacts of lightning and transient luminous events on atmospheric composition, spectroscopic signatures of
ionization, and its potential for X-ray emission in both the laboratory and in lightning.

(d)  Electrical properties of the discharge and their electromagnetic emissions.  This includes coupling to
pulsed power circuits, measurements of inner electric fields, modeling of charge distribution and charge
transport.

(e)  Processes on earlier or later time scales, on the one hand the initiation of streamers, and on the
other hand later processes such as secondary streamers, transition to leaders, glows or arcs, and afterglow
effects, heating and convection.

\section{The papers in the cluster issue and other recent literature}

The articles within the cluster issue are a representative sampling of recent research in various fields in
which streamers play a role.  We here briefly discuss how they relate to each other and indicate where they
fit into the broader field of streamer research, without attempting a complete review.

\subsection{The structure of streamers in ambient air}

The first five articles in the cluster issue deal with the basic macroscopic characteristics of streamers in
air at standard temperature and pressure; this includes their diameter, speed of propagation, and their
branching into trees of many interacting streamers. The article~\cite{Winands} by Winands {\it et al.}\
presents measurements of many positive or negative streamers emerging together from a wire electrode,
showing that streamers of different polarity largely resemble each other. The authors have reported similar 
observations earlier in~\cite{Winands2006, IEEEHans}. In contrast, Eichwald {\it et
al.}~\cite{Eichwald} and Nudnova {\it et al.}~\cite{Nudnova} present measurements and simulations of single
positive streamers that emerge from a needle electrode; Eichwald for d.c.\ voltage and Nudnova for
sharp voltage pulses similarly to the experiments of Winands. The focus on positive streamers is quite generic 
for the recent streamer literature. The measurements presented by Briels {\it et al.}~\cite{BrielsPM} elucidate
this apparent contradiction. They cover a voltage range of 5 to 95 kV with different power supplies and show
that for low voltages only positive streamers emerge while for higher voltages, negative streamers are
generated as well; with increasing voltage the negative streamers more and more resemble the positive ones;
they both form multiply branched trees.

The article of Nudnova {\it et al.}~\cite{Nudnova} builds on earlier work by Pancheshnyi and 
Starikoskii~\cite{Panch03, Panch04, Panch05PRE, IEEENud}; it focuses on simulation and experimental reconstruction 
of the head structure of positive streamers as revealed in optical emissions from the light-emitting shell
around the streamer head.  This inner head structure with its associated strong field enhancement at the
streamer tip is the most important element of a streamer, and its properties play a pivotal role in the
dynamics, as many simulations show.  Its structure as determined in simulations  is here confirmed in
experimental observations.

Within the voltage range of 5 to 96 kV, Briels {\it et al.}~\cite{BrielsPM} find that the streamer diameter
can vary by a factor of 15 and the propagation speed by a factor of 40; they present detailed measurements
of diameters and velocities as a function of the applied voltage, as well as an empirical fit of the
velocity as a function of the diameter. Experiments are here characterized by voltages rather than by
hypothetical average electric fields, as the fields in a wire-to-plane or needle-to-plane electrode geometry
are quite inhomogeneous.  The electric field near the needle or wire electrode where the streamers are
launched is primarily determined by electrode geometry and applied voltage, and depends only weakly on the
gap distance. This local field together with the availability of free electrons and with emission processes 
on the electrode surface determine the streamer inception; this inception process was recently imaged
in~\cite{BrielsInc}. After inception in the high field zone near the electrode, the streamers propagate into
an undervolted region, i.e., into a region where the local electric field in the absence of streamers is 
below the breakdown value.

The asymmetry between positive and negative streamers in air (that also plays in lightning 
physics~\cite{Williams06/PSST}) is studied in the simulations by Luque {\it et
al.}~\cite{Luque}. Here, streamers both in homogeneous and in inhomogeneous fields are studied.
First double headed streamers in homogeneous overvoltage fields are investigated,
similarly to earlier simulations by Liu {\it et al.}~\cite{Liu04,Pasko07}.  Then positive and negative
streamers are launched from the high field region near needle electrodes, and they propagate into a low
field region as in the experiments~\cite{BrielsPM}. Positive streamers propagate with a similar relation
between velocity and radius as in the experiments~\cite{BrielsPM}, while negative streamers at low voltages
broaden and extinguish, also in agreement with experiments.

A characteristic of many streamers and sprites is that discharge channels appear in groups; they can branch
or extinguish, and sometimes they reconnect. The full three-dimensional structure of streamer trees emerging
from a needle electrode was recently analyzed through stereographic imaging, and the branching angles were
determined by Nijdam {\it et al.}~\cite{Nijdam}. Fully three-dimensional simulations of two interacting
streamers were performed by Luque {\it et al.}~\cite{LuquePRL08} showing that equally charged streamer heads
do not necessarily repel each other, but in air they also can merge due to the nonlocal photo-ionization
reaction.

\subsection{The influence of magnetic fields and the role of medium and density}

Manders {\it et al.}~\cite{Manders} study short positive and negative streamers in ambient air in magnetic
fields of 2.5 to 12.5 Tesla.  They observe that the streamer path bends due to the Hall effect and they
reconstruct the local electric fields. These measurements are also relevant for sprite discharges above
75-85 km where the thermal electron collision frequency drops below the cyclotron frequency in the
geomagnetic field.

Streamers are observed not just in air, but in almost any ionizable, nonconducting medium, be it gas, liquid
or solid. We recall that phenomenological concepts of streamer propagation as presented in~\cite{Raizer}
were originally developed in the context of ionization fronts in semiconductors~\cite{Dya}. In the present
cluster issue, the review by Kolb {\it et al.}~\cite{Kolb} on streamers in water and other dielectric
liquids is included. Without aiming at any completeness of references, we also recall investigations in
liquid nitrogen and helium~\cite{Denat1, Denat2} and in gaseous nitrogen-oxygen mixtures~\cite{Yi2002, BrielsN,
BrielsSim}. Discharges in other gases besides air are of interest; we refer to~\cite{Lamps1, Lamps2, Lamps3} 
for streamer processes in the ignition of high pressure discharge lamps, and to the collection of 
articles~\cite{Planets} and to~\cite{Planets2, RD08} for a discussion on discharges in the atmospheres of 
other planets. Images of streamers in ambient air at 1 to 9 bar can be found in~\cite{IEEEHighP}.

Similarity laws between phenomena in different gas densities such as streamers in ambient air and sprite 
discharges at high altitudes of the atmosphere are based on theoretical considerations~\cite{Pasko07, 
Liu06, Ebert06}. These similarity laws are reviewed and experimentally tested on
positive streamers in~\cite{BrielsSim} where the pressure was varied from 0.013 to 1 bar and the gap
distance from 10 to 160 mm; the extrapolation to sprites works very well. We remark that the relevant scales
of electric fields, electron and ion densities, lengths and times depend on gas density, whereas the voltage
does not~\cite{BrielsSim, Ebert06}. The paper by Stenbaek-Nielsen and
McHarg~\cite{Nielsen} on high time resolution imaging of sprite discharges underscores the similarity of
streamers and sprites from the geophysical side; the authors show the same light emission structure of the
head, and analyze onset, velocities, brightness and size and branching, as well as later luminous and bead
structures.

The essence of the streamer mechanism is the self generated field enhancement at the streamer head. Adachi
{\it et al.}~\cite{Adachi} measure approximately the local field in sprite heads through a spectroscopic
method.  They estimate the charge moment transfer of the generating lightning stroke in the underlying
thunderstorm, and deduce from it the evolution of the background electric field in the sprite region. This
work further supports that indeed the streamer mechanism is active in sprite discharges. An important
counterpart to the estimation of local electric fields from observations is the measurement of lightning
currents and of the associated charge moment transfers between cloud and ground. Their global distributions
and seasonal variations are discussed by Sato {\it et al.}~\cite{Sato}.

The nonlocal photoionization reaction is an important ingredient for positive streamers in air specifically,
but the actual reaction rates and photoabsorption lengths are not very reliably known, as discussed
in~\cite{Nudnova,LuquePRL08}. (The different case of repetitive streamer discharges with memory effects is
discussed by Pancheshnyi in~\cite{Panch05}; the d.c. driven streamers studied by Eichwald {\it et
al.}~\cite{Eichwald} are repetitive as well.) There are actually two macroscopic observations that can give
an indirect handle on the role of photo-ionization or pre-ionization relative to impact ionization, namely
the bending of the streamer in a strong magnetic field~\cite{Manders}, and the merging or repulsion of two
streamer heads~\cite{LuquePRL08}. The fact that photoionization is increasingly suppressed at pressures
above approximately 80 mbar and for a decreasing oxygen-nitrogen ratio can be used in  experiments
like~\cite{BrielsPM, Nijdam, BrielsN, BrielsSim} to study this interaction further on macroscopic phenomena.

\subsection{UV and X-ray emissions, chemical products and energy deposition}

A characteristic and important feature of streamer discharges and transient luminous events are UV and X-ray
emissions, chemical products and energy deposition.

X-ray emissions and gamma ray bursts from lightning are currently under very active investigation; the
cluster issue contains two papers that investigate possible sources of intensive radiation. Nguyen {\it et
al.}~\cite{Nguyen} measure multiple gamma ray bursts from a MV discharge in the laboratory. In contrast to
earlier experimental investigations~\cite{Dwyer,Rahman}, they clearly attribute the bursts to the
streamer-leader phase of the discharge. Milikh {\it et al.}~\cite{Milikh} propose that UV flashes observed
from a satellite are due to the streamer zone of upward propagating gigantic blue jets. Theoretical
investigations~\cite{Moss, Li07, Chanrion08, Li08a, Li08b} of this effect are reviewed in section~\ref{sim}.

The total energy of sprites and other transient luminous events as inferred from three years of optical
emission measurements made with the ISUAL instrument aboard the FORMOSAT2 satellite are described Kuo {\it
et al.}~\cite{Kuo}.  While the energy deposition per event is of the order of 50 MJ for sprites, halos and
elves, most of the energy is deposited in the upper atmosphere by elves, as they are the most frequent.

The streamer head can be considered as a self-organized, highly efficient plasma reactor that travels
through the gas; it initiates various chemical reactions that are important both in technical applications
and in atmospheric chemistry. Van Heesch {\it et al.}~\cite{Heesch} report extremely efficient
generation of O$^*$ radicals and ozone with rapidly pulsed streamer corona discharges when fast voltage 
pulses in the range of 50 to 90 kV are applied; they even reached a conversion of dissipated electrical 
energy into ozone above 50\%. 
O$^*$ radicals and ozone are widely used for disinfection purposes; its generation in other 
corona reactors is discussed in~\cite{Winands2006, Eichwald, OnoOda}.

Sprite chemistry is a new subject treated in the cluster issue by Gordillo-Vazquez~\cite{Vazquez} following
the seminal study of Sentman {\it et al.}~\cite{SentmanChem}.  These studies model the nonstationary,
non-LTE chemical response of the upper atmosphere near ~70 km to the passage of a streamer of typically
observed magnitude.  Using a chemical kinetics scheme involving several dozen neutral and ionic species and
several hundred coupled chemical reactions, the models follow the production, loss and chemical
transformation of species through a complex sequence of processes starting with breakdown in the streamer
head and continuing until the system returns to predischarge equilibrium.  The models include what are
believed to be the principal processes involved in electrical discharges; ionization, excitation,
dissociation, collisional and radiative deactivation, attachment, detachment, recombination, charge
exchange, and various chemical reactions.

\subsection{Simulations and theory}\label{sim}

Due to the many length and time scales involved and due to their nonlinear coupling in a dynamics far from
equilibrium, streamer physics is a challenging multiscale problem~\cite{Ebert06}. The simulation of
propagating single streamer in cylindrical symmetry in fluid approximation with its inner shell structure is
after more than 20 years of study now reaching maturity, a few recent papers are~\cite{Eichwald, Nudnova, 
Panch03, Panch04, Panch05PRE, Luque, Liu04, Pasko07, Liu06, Panch05, MontijnJCP06, MontijnJPD06, Bourdon07, 
Liu07, Luque07}. The cluster issue only contains such studies, plus a study by Naidis~\cite{Naidis} of 
repetitive discharges and the transition from streamer to the spark break-down; these long times can 
presently only be treated by analytical approximations. In the present section we also briefly review other 
recent developments in simulations and theory.

An efficient computational treatment of the nonlocal photoionization reaction in oxygen-nitrogen mixtures
like air for fluid models was recently proposed independently by S\'egur {\it et al.}~\cite{Bourdon07, Liu07,
Bourdon06} and Luque {\it et al.}~\cite{Luque07}; the cluster issue contains a detailed follow-up
study of different computational strategies for calculating the monochromatic radiative transfer in air in
fluid models for streamers~\cite{Capeillere}.

A vital computational issue is adaptive grid refinement that is capable of accurately resolving the inner
spatial structure of the streamer head while avoiding the high computational costs of solving Laplace's
equation in the large outer non-ionized region on the same fine grid as the in head region.  Adaptive grid
refinement in finite volume codes for single streamers was developed and thoroughly discussed by Montijn
{\it et al.}~\cite{MontijnJCP06}, it was extended by Luque {\it et al.} with photo-ionization~\cite{Luque07}
and to full three dimensions~\cite{LuquePRL08}. The cluster issue contains a study on adaptive grid refinement 
in finite element codes
by Papadakis {\it et al.}~\cite{Papadakis}; another parallel adaptive mesh refinement code on Cartesian grids 
was recently presented by Pancheshnyi {\it et al.}~\cite{Panch08}; Kolobov {\it et 
al.} also have recently published some streamer results based on dynamically adaptive Cartesian 
meshes~\cite{IEEEKolobov}. Note that
avalanches~\cite{MontijnJPD06}, propagating streamers and unstable branching streamers form a ladder of
increasing computational complexity as density gradients become steeper and the physical branching
instability should be clearly distinguished from a numerical instability. The branching time was found to
saturate with decreasing size of the numerical grid in simulations by Montijn {\it et
al.}~\cite{MontijnPRE06}; these simulations were performed in cylindrical symmetry as many before and
therefore in fact give an upper bound for the branching time in full three dimensions, but they do not
follow the true evolution after branching.

Full three-dimensional simulations of streamers in fluid approximation pose a major computational
challenge. Interesting phenomena include the full branching structure after the streamer head has
destabilized, the interaction of several streamers and almost all interactions of streamers with 
electrodes, walls, particles or local inhomogeneities. Kulikovsky~\cite{Kuli} and 
Pancheshnyi~\cite{Panch05} have presented images of 3D simulations of streamers at the moment
of branching with low numerical accuracy on uniform grids. Luque {\it et al.}~\cite{LuquePRL08} 
recently have developed a parallelizable 3D code with dynamically adapting grid refinement, and 
they have studied the interaction of two streamers with this method.

The study of interfacial motion in nonlinear dynamics gives firm theoretical support for the statement that
streamers can branch without any noise or disorder if the streamer head has developed into a dynamically
unstable state. In Section 5 of~\cite{Ebert06}, this branching concept of a Laplacian instability is 
confronted with the old branching concept based on stochastic ionization avalanches that can be found in 
many textbooks and that can be traced back to Raether in the 1930's.  An extensive
analysis of a moving boundary approximation for the deterministic fluid model in~\cite{Meul04, Meul05, 
Scha07, BrauPRE08} places streamer branching in the mathematical context of other branching problems in 
nature such as viscous fingering, dendritic solidification etc.

Even though streamers can branch in fully deterministic fluid models, random motion of individual particles
or a disordered background can trigger an inherent branching instability earlier than it would occur in a
deterministic system. The influence of random initial conditions was studied in a 1D model by Array\'as {\it
et al.}~\cite{Manuel08}. The influence of "bubbles" of slightly different density on streamer branching in
2D was recently studied by Babaeva and Kushner~\cite{BabKush08}, while they studied the influence of dust
particles on streamer propagation in~\cite{BabKush06}. These studies are obviously relevant for discharges
in liquids with bubbles~\cite{Kolb} as well as in thunderclouds or in dust devils on Mars; these processes
as well as sliding discharges along dielectric walls will not be further reviewed here.

There are hundreds to tens-of-thousands of streamer heads in the laboratory~\cite{Winands,Heesch}, or in the
streamer zones of lightning leaders and of blue jets or in sprite discharges; to model them in fluid
approximation is beyond the capabilities of the current generation of computers, therefore models must be
reduced.  Several simple electrodynamic approximations have been proposed, e.g., by Bazelyan and
Raizer~\cite{Raizer}, and they are extended and used in~\cite{Milikh}. 
In a search for such simple electrodynamic models,
the heads of simulated single streamers are characterized by charge content, voltage, radius and velocity
in~\cite{Luque}. On the other hand, the study~\cite{ST08} shows that a group of streamers can exhibit
different electrodynamic behavior than a single streamer.  In particular, the electric field in the streamer
channels can be completely screened in a group of streamers, but not in a single streamer.

On the other end of the hierarchy of length scales, the stochastics of single particles may play a role in
the development and dynamics of streamers; ultimately, a fluid model is just an approximation of the true
particle dynamics. In fact, a comparison of a Monte Carlo and a fluid realization of the same streamer
ionization front shows that electron energies and the plasma density behind the front increasingly deviate
with increasing electric field~\cite{Li07}. This is, of course, a pre-cursor of the electron run-away effect
that could be a source of X-ray or gamma radiation of energetic streamers. This idea motivated Moss {\it et
al.}~\cite{Moss}, Li {\it et al.}~\cite{Li07} and Chanrion and Neubert~\cite{Chanrion08} to investigate
Monte Carlo models of streamers and sprites. Moss and Li did this in one-dimensional approximation while
Chanrion studies three-dimensional streamers and uses so-called superparticles. However, Li {\it et al.}\
have demonstrated in another paper~\cite{Li08a} that superparticle simulations of avalanches and streamers
can suffer from strong numerical artifacts.  They are developing a hybrid model~\cite{Li08b} that treats the
high field region with low electron numbers using a Monte Carlo model, and the ionized streamer interior
with a fluid model. Monte Carlo models or hybrid Monte Carlo-fluid models can also track the primary
excitations of molecules after the passage of the streamer front as an input for chemistry models, as well
as trace fluctuations of particle numbers as a possible additional trigger for the streamer branching
instability.

\section{Acknowledgements}

We thank the staff of the Lorentz Center in Leiden, The Netherlands, for the administrative support of the
workshop, in particular, Wies Groeneboer, Henriette Jensenius, Maartje Kruk, and Wim van Saarloos. We
acknowledge financial support for the workshop by the Lorentz Center, by the Dutch research school "Center
for Plasma Physics and Radiation Technology (CPS)", by the group "Elementary Processes in Plasmas (EPG)" at
Eindhoven University of Technology, by the Netherlands' Organization for Scientific Research (NWO) through
"Exacte Wetenschappen (EW)", through the national research school "Nonlinear Dynamics of Natural Systems
(NDNS)", and through a joint Dutch-Russian research grant, and by the companies Philips and Renault. We
thank the editors of J.~Phys.~D for the invitation to be guest editors of the present cluster issue. And
finally, we thank the staff of IOP, in particular, Lesley Fifield, for the careful handling of the
manuscripts.

\end{document}